# Physics-Constrained Neural Network for Metasurface Optical Response Prediction

Reza Masoudian Saadabad[1*], Ramin Emadi[1], Lingraj Kumar[2], Davide Bacco[2], and Maja Colautti[1]

[1] National Institute of Optics (CNR-INO), Via Nello Carrara 1, Sesto Fiorentino 50019, Italy
[2] Dipartimento di Fisica e Astronomia, Università degli Studi di Firenze, Via G. Sansone 1, Sesto Fiorentino 50019, Italy

* Corresponding author: reza.masoudiansaadabad@ino.cnr.it

**Abstract:** A physics-constrained neural network is presented for predicting the optical response of metasurfaces. Our approach incorporates physical laws directly into the neural network architecture and loss function, addressing critical challenges in the modeling of metasurfaces. Unlike methods that require specialized weighting strategies or separate architectural branches to handle different data regimes and phase wrapping discontinuities, this unified approach effectively addresses phase discontinuities, energy conservation constraints, and complex gap-dependent behavior. We implement sine-cosine phase representation with Euclidean normalization as a non-trainable layer within the network, enabling the model to account for the periodic nature of phase while enforcing the mathematical constraint $sin^2\varphi + cos^2\varphi = 1$. A Euclidean distance-based loss function in the sine-cosine space ensures a physically meaningful error metric while preventing discontinuity issues. The model achieves good, consistent performance (e.g., coefficient of determinations above 0.9) with small, imbalanced datasets of 580 and 1075 data points, compared to several thousand typically required by alternative approaches. This physics-informed approach preserves physical interpretability while reducing reliance on large datasets and could be extended to systems involving periodic or wrapped quantities.

**Keywords:** Physics-informed neural networks, Machine learning, Nanophotonics, Metasurface, Optical response

## 1. Introduction

The interaction of light and matter at the nanoscale is key in governing the performance and design of nanophotonic devices for the manipulation of electromagnetic waves in a broad range of applications spanning from photodetection [1] and optical imaging [2], to telecommunication [3] and quantum photonic technologies [4,5]. Among nanophotonic structures, metasurfaces consisting of arrays of subwavelength-scale antennas have attracted great interest across numerous domains for their ability to control light-matter interactions. They enable the efficient manipulation of light characteristics such as phase, amplitude and polarization [6–17] , while offering the advantage of a flat design ideal for compactness and on-chip integration. Some examples of applications enabled by metasurface-tailored interactions include flat lenses (metalenses) [18], optical beam steering [19], holography [20], sensing, spectroscopy and quantum optical computing [13], and reconfigurable optics [21]. Traditionally, metasurface design relies on forward predictions of the optical properties through Finite Element Methods (FEM) and Finite Difference Time Domain (FDTD) methods. While accurate, these approaches are computationally expensive, and the optimization process lacks a fully systematic framework, particularly when considering the multi-parameter problems [14] due to the collective nature of their effects. Thus, optimizing metasurfaces with forward prediction methods is inherently challenging.

Therefore, recent efforts have been incorporating machine learning approaches to metasurface design to achieve a more systematic optimization [22–24], enabling to reduce computational time significantly while predicting solutions for new design problems. Machine learning algorithms can indeed enable the systematic exploration of parameters by learning sophisticated relationships between the structural characteristics of metasurfaces and their optical properties through training datasets provided by standard simulations for specific geometries. It has been demonstrated that deep learning is more efficient than conventional optimization methods to predicting optical response of simple metasurfaces [25]. Machine learning has also been used for phase response optimization using binary-coded metasurfaces [26], as well as to design multifunctional metasurfaces [27]. However, standard machine learning methods, being purely data-driven, may struggle with fundamental physical constraints of optical systems, such as energy conservation and phase continuity, and predict non-unique solutions for the same input configuration. Physics-informed neural networks (PINNs) have emerged in recent years to overcome such challenges by incorporating physical laws directly into the training process [28–33], allowing to increase the accuracy of the predictions and rely on smaller datasets for the training process [14]. As an example, by incorporating the Helmholtz equation into the neural network PINNs are used to optimize the design of a single cylinder that replicates the electric field response of an array of cylinders, showcasing the PINNs performance for efficient design optimization [29]. Recently, PINNs have also been used to derive the resonant frequency of structures, enabling more accurate modeling of the time dynamics of optical resonances [34].

In this work, we propose a physics-constrained neural network specifically designed to incorporate physical laws in the prediction of metasurface optical response. While previous physics-informed methods have incorporated physical laws at the differential equation level (e.g., Maxwell's equations or reciprocity) [33,35–37], our approach introduces circle normalization and energy conservation as a physics-informed non-trainable layer in the network. This strategy effectively addresses the complex dependence of optical phase behaviour on the subwavelength-scale gap (i.e., the distance between adjacent nanostructures), a key challenge in metasurface design. Previous studies have employed various approaches to address phase discontinuities. For instance, Jiang et al. [38] attempted direct phase prediction despite the challenges posed by discontinuities at the 0/2 $\pi$ boundaries, while An et al. [39] used two separate neural networks to predict the real and imaginary components of the transmission coefficient, from which phase and amplitude were derived. In contrast, our method directly incorporates the circular nature of phase into a single, unified network, eliminating the need for specialized weighting strategies for imbalanced datasets or separate network branches to handle different data regimes [40–42]. This approach ensures physically meaningful outputs across all regimes, reduces network complexity, and avoids discontinuities. High accuracy is achieved with relatively small, imbalanced datasets – ranging from 290 to 1075 samples - compared to the several thousand typically required in prior works [38,43]. The systematic evaluation across three dataset sizes (290, 580, and 1075 data points) demonstrates the robustness of the physics-constrained approach while confirming the natural relationship between dataset size and model performance. Moreover, the physics-informed loss function provides a consistent and physically meaningful error metric to assess the network performances.

## 2. Physics-Constrained Neural Network Framework

### 2. 1 Dataset Characteristics

In this work, we explore the optical response of a metasurface composed of a homogeneous 2D array with a single nanodisk in the unit cell (refer to Supplementary Material to see metasurface configuration and simulation details). We use three training datasets, obtained by FEM simulations using JCMsuite, to evaluate performance across different data scales: Dataset A with 1075 data points, Dataset B being a subset of A with 580 data points, and Dataset C being a further subset with 290 data points. Each data point represents a unique configuration described by the metasurface structural parameters, such as diameter and height of the nanodisk, period, and the incident angle of the incoming light, with the resulting phase and amplitude of transmitted light. The FEM simulation models a single unit cell, with symmetric boundary conditions applied to simulate the response of an infinite nanodisk array. These datasets are not only small but also have imbalanced distributions (Figure I-a illustrates this for the gap parameter). This combination makes it challenging to develop a high-performing model, providing a strict test for our approach. Dataset C, with only 290 data points, represents an extreme case of data scarcity. Indeed, standard neural network typically need large, balanced datasets to achieve good generalization [41,44], i.e., the model's ability to accurately predict new, unseen data with different structural parameters and optical properties.

Figure I-b shows the phase distribution across different gap sizes in Dataset A, and Kernel Density Estimation (KDE) providing a smoothed view, overlaid with scatter points representing the phase values for each gap size. The gap refers to the physical distance between adjacent nanodisks, calculated by subtracting diameter from period. The KDE visualization reveals that for the large-gap regime (approximately > 50 nm), phase values tend to cluster mainly around the range of 2-4 radians. In contrast, in the small-gap regime (approximately below 50 nm), metasurface exhibits a volatile phase change with values scattered across the entire 0-2π range, a behaviour which is physically expected due to stronger near-field coupling between adjacent structures and which causes a significant learning challenge for neural networks due to the high sensitivity to even tiny gap variations. Samples from this region are critical to explore the full potential of metasurfaces, as they are often related to high-Q resonances and enhanced light-matter interactions [45–53], making them essential for evaluating the network's performance in this regime.

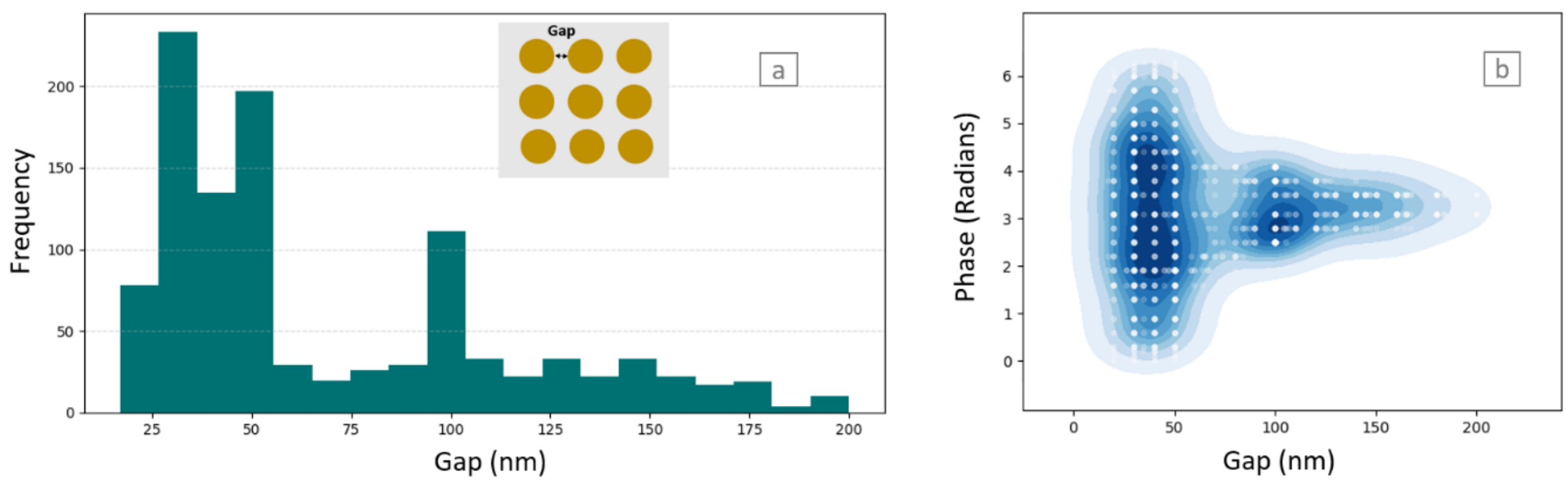

Figure I: a. Histogram shows the frequency of gap values within Dataset A. As shown in the inset, gap refers to the distance between adjacent nanocylinders in metasurface. b. Kernel Density Estimation (KDE) with scatter points illustrates the distribution of phase values across different gap sizes in Dataset A. The darker blue regions represent higher concentrations of data points.

A further challenge is addressing phase discontinuities. Standard machine learning models treat the 0 and $2\pi$ boundary points as distant values rather than the same physical states, creating artificial errors that compromise prediction accuracy. To address this, we formulate this problem as a physics-constrained regression task that predicts the amplitude and phase of transmitted light from the metasurface structural parameters. Our neural network efficiently handles both previously identified gap regimes while respecting the circular nature of phase values. To ensure physical consistency, the model bounds the amplitude values to [0 , 1] since transmittance cannot exceed 100%, and enforces the Pythagorean identity for phase, $sin^2\varphi + cos^2\varphi = 1$, ensuring that the sine and cosine components of each predicted phase angle remain physically valid. In the following paragraphs we describe the details of the neural network architecture implementations.

**2.2 Overall Architecture**

As shown in Figure II, our neural network accepts input parameters (e.g., diameter and height of the nanodisk, period, refractive index, and the incident angle of the incoming light) that represent the material and geometric properties of the metasurface unit cell. It follows a structure with multiple dense (*i.e.* fully connected) layers, starting with a wide layer of 256 neurons that gradually decreases in width (layers of 128, and 64 neurons) to extract relevant information. The network is then divided into two specialized output paths: one for amplitude, and one for phase, where each branch includes a dense layer of 32 neurons serving as feature extractors for each output. To enhance learning and ensure balanced features propagation through the network, batch normalization followed by ReLU (Rectified Linear Unit) activation are applied to all dense layers. Batch normalization standardizes the input to each layer, promoting training stability, while ReLU introduces non-linearity, enabling the network to learn complex patterns. To reduce the risk of overfitting - when a model learns the training data too closely and performs poorly on new data - we also include dropout layers (with a dropout rate of 0.1) after the first two, randomly disabling a fraction of neurons during training to encourage more robust learning.

We used different activation functions for the branches based on their underlying physics:

1- Amplitude branch uses a sigmoid activation function constraining outputs to the physically meaningful range of [0 , 1] – considering amplitude normalization.

2- Phase branch uses a *tanh* (hyperbolic tangent) activation function constraining outputs to [-1 , 1], handling phase wrapping using sine and cosine components representation.

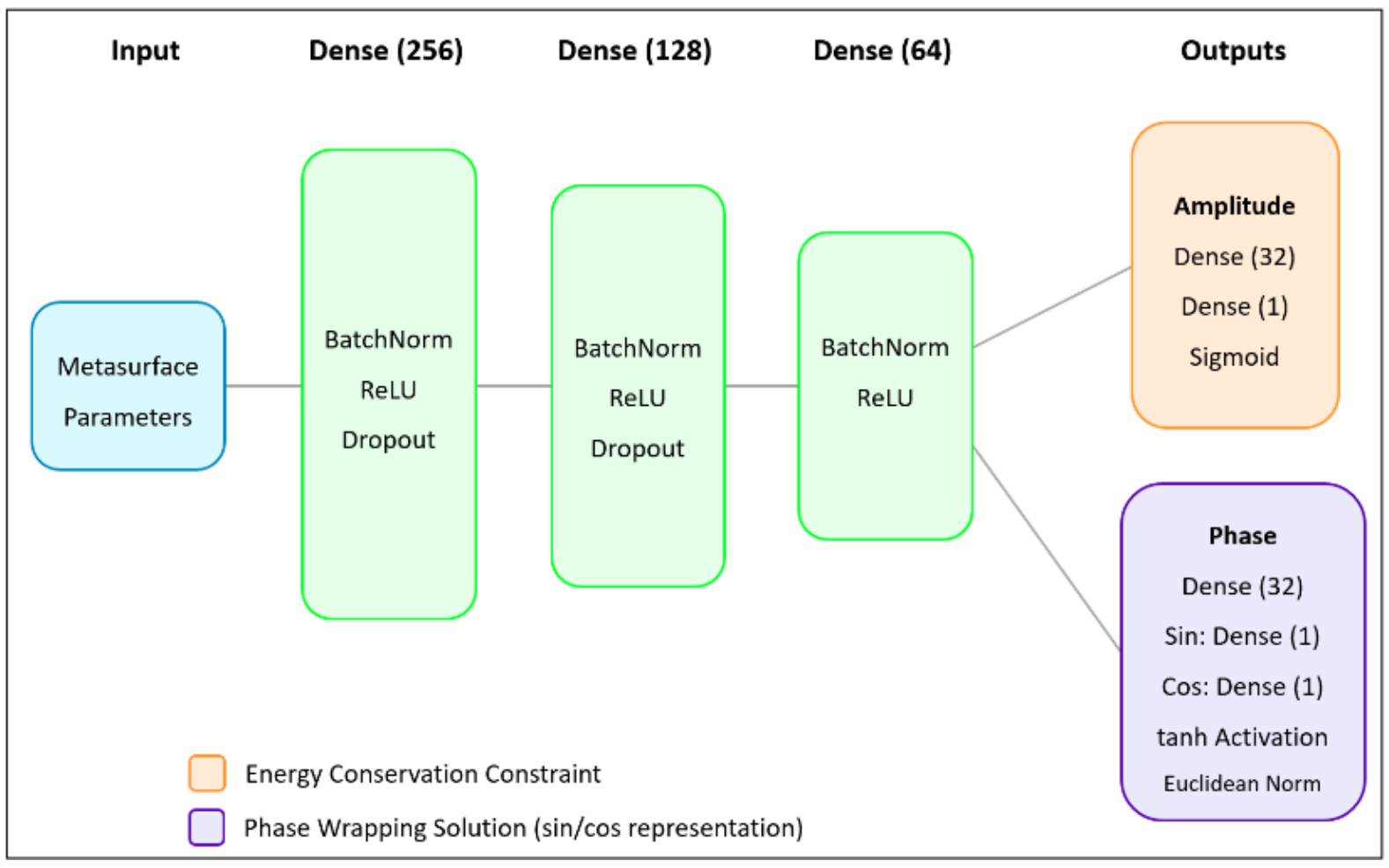

Figure II: Graphical representation of the physics-constrained neural network.

In particular, the specific physical constraints embedded in the network and loss function are discussed below.

**Sine - Cosine Representation for Phase Wrapping**

To address the phase wrapping issue, the network's phase branch is decomposed into two output nodes: the sine and cosine components of the phase angle $\varphi$, rather than predicting $\varphi$ directly. Both components are predicted simultaneously using hyperbolic tangent (*tanh)* activation functions, which naturally constrains outputs to the [-1 , 1] range, consistent with the behaviour of sine and cosine. This representation offers clear advantages over direct phase prediction. It inherently avoids discontinuities at the $0/2\pi$ boundary, since equivalent phase values like $\varphi$ and $\varphi + 2\pi$ are mapped to the same point. Thus, the network does not need to learn periodic equivalences, they are built into the formulation. Furthermore, this approach prevents misleading loss function (difference between predicted and real value as from simulation) gradient signals near phase boundaries. For example, a true phase of 0.1 radians and a predicted phase of $2\pi - 0.1$ would appear numerically distant in direct regression, despite being nearly identical physically. The sine-cosine representation ensures smooth loss behaviour in such cases, and avoids complex unwrapping algorithms [54].

**Euclidean Normalization for Phase Components**

The fundamental identity $sin^2\varphi + cos^2\varphi = 1$ is enforced for any valid phase angle $\varphi$ via the Euclidean normalization of the model's raw sine and cosine outputs components:

$$\sin\hat{\varphi} = \frac{\sin\varphi}{\sqrt{\sin^2\varphi + \cos^2\varphi}}$$
$$\cos\hat{\varphi} = \frac{\cos\varphi}{\sqrt{\sin^2\varphi + \cos^2\varphi}} \tag{1}$$

As shown in Figure II, this normalization is implemented directly within the model architecture as a non-trainable layer rather than applying it after prediction. This design ensures that gradients during backpropagation pass through the constraint, allowing the model to learn physically consistent outputs inherently, rather than adjusting them after training. As a result, the network achieves better generalization even with limited training data.

**Energy Conservation Constraints**

To ensure the model respects energy conservation principle, we bound the amplitude prediction to the [0 , 1] range by a sigmoid activation function. Additionally, we include an energy conservation term in the loss function that penalizes the model if its predictions violate the physical principle. Here below the employed loss functions are detailed.

**Loss Function Formulation**

**I. Amplitude loss**

$$L_{amp} = \frac{1}{N}\sum_{i=1}^{N}(A_{p,i} - A_{t,i})^2 \tag{2}$$

where $A_{p,i}$ and $A_{t,i}$ are the predicted and true amplitude values for the sample *i* respectively, and *N* is the training dataset size.

**II. Energy conservation loss**

$$L_{energy} = \frac{1}{N}\sum_{i=1}^{N}(\max(0, A_{p,i} - 1))^2 \tag{3}$$

This constraint penalizes only predictions surpassing the physical limit of perfect transmission ($A_{p,i}$= 1), ensuring compliance with the energy conservation principle.

**III. Phase loss**

Leveraging sine-cosine representation, the error is calculated based on the Euclidean distance between true and predicted phases on the unit circle in sine/cosine space:

$$L_{phase} = \frac{1}{N}\sum_{i=1}^{N}\sqrt{(\sin\hat{\varphi}_{p,i} - \sin\varphi_{t,i})^2 + (\cos\hat{\varphi}_{p,i} - \cos\varphi_{t,i})^2} \quad (4)$$

This is the chord distance between unit-normalized phase vectors. This metric ranges from 0 (perfect prediction, where true and predicted phase values coincide) to 2 (maximum error, where true and predicted phase values are diametrically opposite on the unit circle). Since cosine and sine terms are dimensionless, the Euclidean distance is also dimensionless. It naturally handles phase wrapping at 0/2π boundaries, since points near these boundaries remain close in sine-cosine space even when they appear distant in conventional phase representation.

**IV. Combined loss**

The final loss function combines all the loss components with adjustable weights ($w_{amp}$, $w_{energy}$, and $w_{phase}$) that represent their relative physical significance:

$$L_{total} = (w_{amp} \times L_{amp}) + (w_{energy} \times L_{energy}) + (w_{phase} \times L_{phase}) \quad (5)$$

These weights are determined empirically, with higher values assigned to loss components corresponding to the most critical physical constraints. The chosen weight values balance the influence of each constraint in the overall loss function, ensuring that no single aspect dominates the training process. This physics-constrained loss function enables the neural network to capture the complex, non-linear relationships between structural parameters and optical response while ensuring physical consistency. The weights reflect the relative importance we assigned to each constraint during optimization. However, we have noticed that the physics-informed architecture itself (i.e., the sigmoid activation for amplitude, sin/cos representation for phase, and unit circle normalization) provides robust constraint enforcement even with lower weight values, demonstrating that the architectural design choices are more critical than loss weight tuning.

## 3. Results and Discussion

Naturally, due to the imbalanced datasets and the presence of regions with different behaviors, we may hypothesize that specialized approaches would be necessary. These could include either specialized weighting strategies [55,56] —adjusting each region's contribution to the loss function— or separate architectural branches for different data regimes. Our analyses, however, revealed that our physics-informed architecture is sufficiently robust to handle both gap regimes effectively with uniform weighting. This suggests that incorporating physical laws directly into the model architecture provides a space where the model can learn different regimes with equal emphasis within a simpler, unified approach.

Our neural network was implemented using the machine learning framework TensorFlow and the programming interface Keras. For all three datasets (A, B, and C), the model randomly splits samples into training (70%), validation (20%), and test (10%). Compared to fixed dataset divisions, this random splitting approach more effectively evaluates the model's ability to generalize rather than merely memorizing specific data points. Please refer to Supplementary Materials to see ML implementation in detail. Figure III shows how the loss components evolution during training for both training and validation datasets. The amplitude loss quickly converges (Figure III-a), indicating that the model immediately learns the amplitude prediction task. It is not a surprising result, because most of amplitude values in the dataset are above 0.9 (due to the material and light wavelength used for the metasurface which fall in a high transparency window), making the amplitude prediction relatively straightforward. Energy loss rapidly converges to zero even in the earliest training epochs (sessions) (see top diagram in Figure III-b), showcasing the advantage of embedding the energy conservation directly into the model architecture through the sigmoid activation in the amplitude branch. In contrast, without this physically-informed activation, the model requires several epochs to learn this constraint (see bottom diagram in Figure III-b). The phase loss, however, converges more gradually, reflecting the complexity of phase prediction (Figure III-c). As a result, the model requires a longer training period, achieving effective convergence over 350 epochs, during which both training and validation losses stabilize around 0.1. The periodic fluctuations observed in the loss curves align with the cosine annealing learning rate scheduler used in our approach. In our model we used annealing learning rate scheduler that is a common technique in machine learning applications to gradually reduce the learning rate during training process in order to improve convergence stability. The trajectories of both curves suggest strong generalization without signs of overfitting (i.e., large difference between train and validation data), indicating that the learning rate modulation effectively guided the model through the loss landscape while preserving performance.

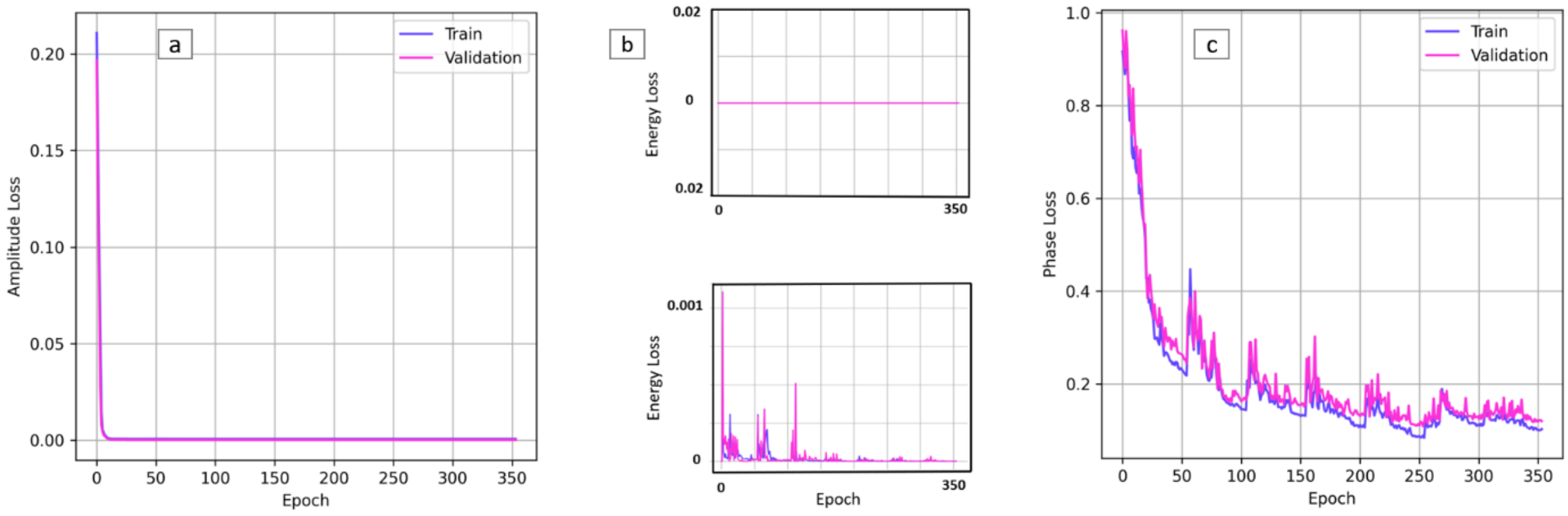

Figure III: Evolution of a. Amplitude, b. Energy, and c. Phase loss components during training. The top and bottom diagrams in b correspond to the models with and without sigmoid activation in the amplitude output, respectively.

Table I presents a comparison among the three datasets in terms of model's performance, averaging over 10 runs with different random initializations for each dataset. The amplitude loss and the energy conservation loss are negligible and effectively zero, respectively, confirming the effectiveness of the physics-constraint architecture. The phase loss, measured in terms of Euclidean distance in sine/cosine space between true and predicted phase values, stabilizes around 0.11 for both datasets A and B, corresponding to a prediction error of about 5.5%. To quantify the phase prediction error in radians, we can calculate the angular difference between true and predicted phase values using the dot product of their unit vector representations. For each data point, we calculate $\cos(\varphi_t)\,.\cos(\hat{\varphi}_p) + \sin(\varphi_t)\,.\sin(\hat{\varphi}_p)$, which yields the cosine of the angular difference. We then apply the inverse cosine function to obtain the angular difference in radians. This approach provides physically meaningful error in radians, which is interpretable in the context of optical phase. The resulting mean angular error for Dataset A is 0.114 radians (approximately 6.5°), while for Dataset B, it is 0.124 radians (approximately 7.1°), with $\mathrm{R}^2$ (known as coefficient of determination) being 0.928 and 0.906, respectively - where 1 identifies perfect performance. For a practical metalens example (100 µm diameter, 100 µm focal length, 780 nm operational wavelength), these phase prediction errors result in focal point displacements of approximately 0.134 µm (Dataset A) and 0.146 µm (Dataset B) at the lens edge – only minor deviations from the ideal.

Dataset C with 290 data points represents an extreme data shortage scenario to test the robustness of our physics-constrained approach under severe data limitations. As expected, Dataset C demonstrates performance degradation with phase loss increasing to 0.161 and $\mathrm{R}^2$ dropping to 0.859, while amplitude and energy conservation constraints remain robust. Notably, the substantially larger standard deviations (±0.056 for phase loss, ±0.130 for $\mathrm{R}^2$) indicate increased variability and less reliable performance across different runs. This confirms the natural relationship between dataset size and model performance while showing that even with only 290 data points, the physics-constrained network can still provide predictive capability, albeit with increased volatility in predictions.

Figure IV provides insight into the angular error distribution for Dataset A and Dataset B. The figure shows the angular error from one representative run for each dataset. Both histograms exhibit a positive skew, indicating that most predicted phase values have low error. Across all 10 runs, the majority of errors remain below 0.2 radians for both datasets.

The model demonstrates consistent performance across both Datasets A and B, with a slight improvement in performance metrics and a reduction in standard deviations when using the larger dataset. It indicates that the physics-constrained approach maintains robust performance even with the smaller Dataset B. The model's strong performance with limited data is especially beneficial for metasurfaces design, where generating training samples via full electromagnetic simulations is both computationally time-consuming and expensive.

| | Amplitude Loss | Energy Loss | Phase Loss | Angular Error (Radians) | $R^2$ (phase) |
|---|---|---|---|---|---|
| Dataset A | 0.001 ± 0.001 | 0.000 ± 0.000 | 0.107 ± 0.027 | 0.114 ± 0.033 | 0.928 ± 0.051 |
| Dataset B | 0.001 ± 0.001 | 0.000 ± 0.000 | 0.113 ± 0.031 | 0.124 ± 0.043 | 0.906 ± 0.074 |
| Dataset C | 0.001 ± 0.002 | 0.000 ± 0.000 | 0.161 ± 0.056 | 0.172 ± 0.065 | 0.859 ± 0.130 |

Table I: Performance metrics of the physics-constrained neural network (mean ± standard deviation over 10 runs).

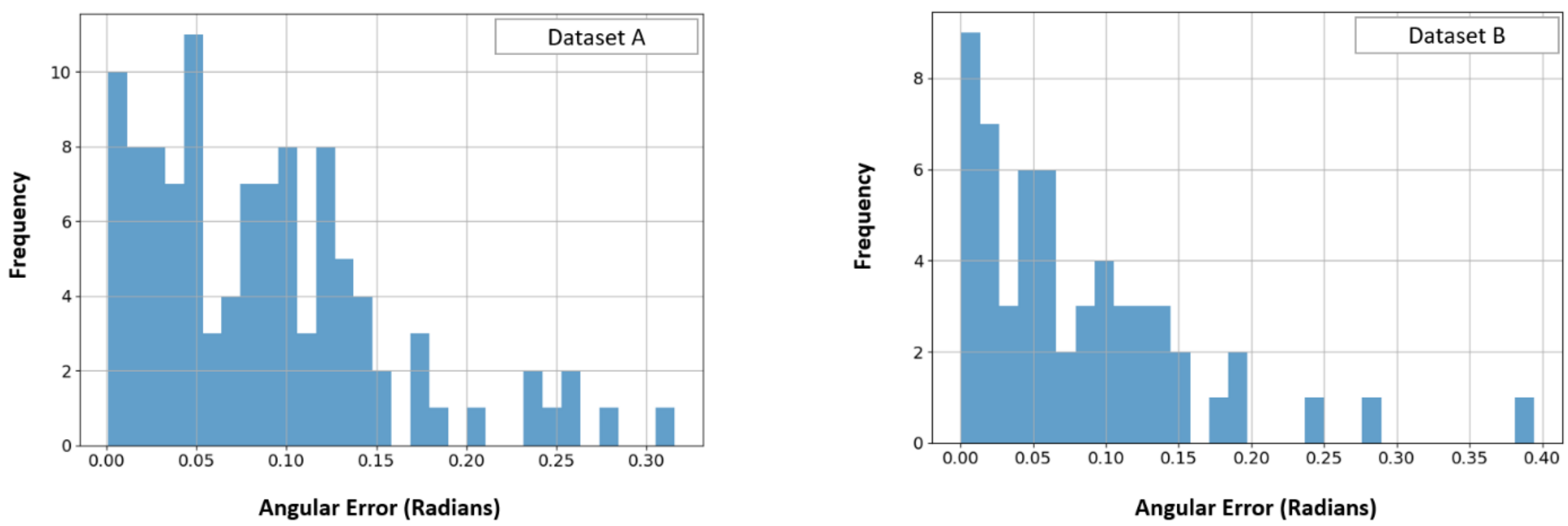

Figure IV: Angular error distribution: the y-axis indicates the frequency of error occurrences.

Figure V highlights the model's prediction capabilities from two perspectives: the relationship between gap size and phase response and the correlation between true and predicted phase values. Figure V-a demonstrates how the model captures the nonlinear relationship between gap size and phase response. The scatter plot shows that the physics-constrained network successfully models this relationship across entire range. Figure V-b offers a direct comparison of the true and predicted phase values, with a color scale indicating the prediction error in sine/cosine space. The strong diagonal alignment along the perfect-prediction line (red dashed line) confirms a high $R^2$ value. An example of the model handling phase wrapping is shown at the black arrow, where the true phase is 0 radian, while the predicted phase is near $2\pi$. Despite the discrepancy, the low error indicated by the color scale reflects the model's ability to recognize the two states as physically close, highlighting the advantage of the physics-informed approach in managing the circular nature of phase values.

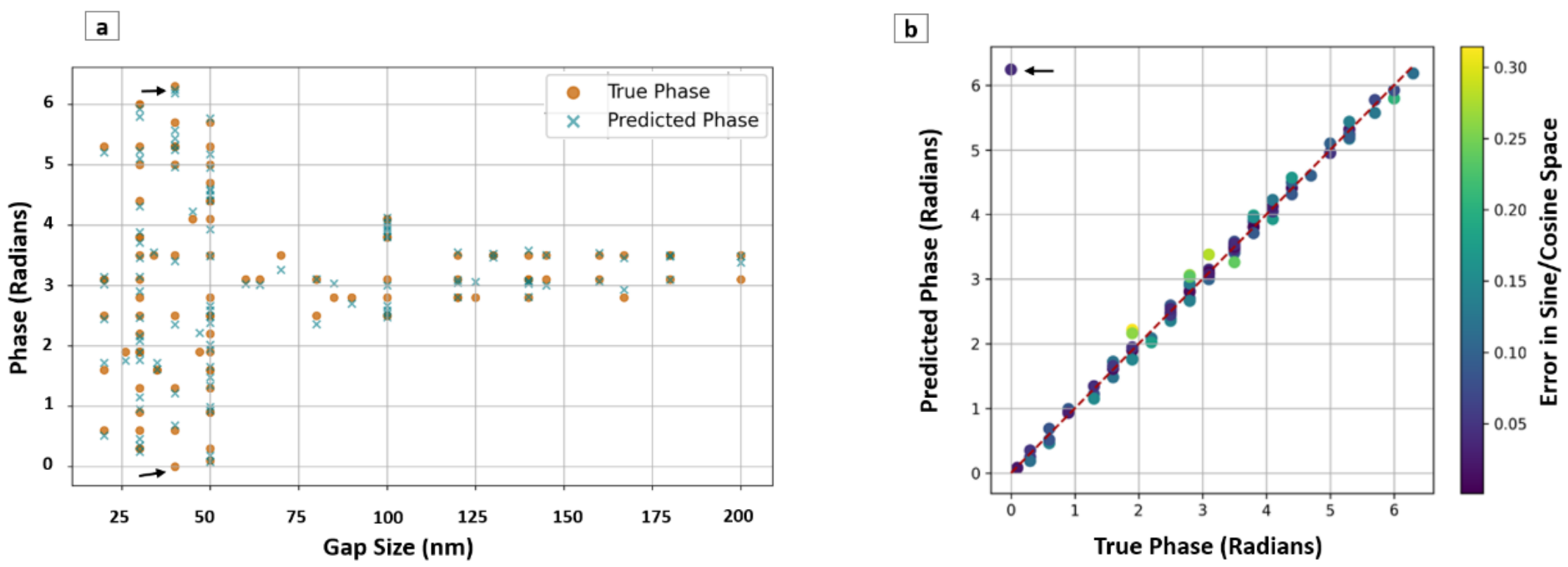

Figure V: a. Predicted and true phase values across different gap sizes. Points that appear as single markers may represent multiple overlapping data points with similar values. b. Direct comparison between the true and predicted values of phase. Color scale indicates prediction error in sine/cosine space. The perfect-prediction is shown by red dashed line. Black arrows highlight a data point where true phase is 0, while the prediction is near $2\pi$. Despite the large numerical difference between these values, the color indicates a low error.

Figure VI evaluates the impact of Euclidean normalization by comparing phase components distributions for the neural network's predictions with and without this constraint. In Figure VI-a, the predicted values (orange dots) deviated from the unit circle (dashed grey), indicating physically impossible phase values that violate the cosine-sine relationship and introduce artifacts in applications. In contrast, Figure VI-b shows predictions with Euclidean normalization, where phase components tightly constrained to the unit circle. This constraint reduces the amount the model needs to learn from data, improving generalization with less data.

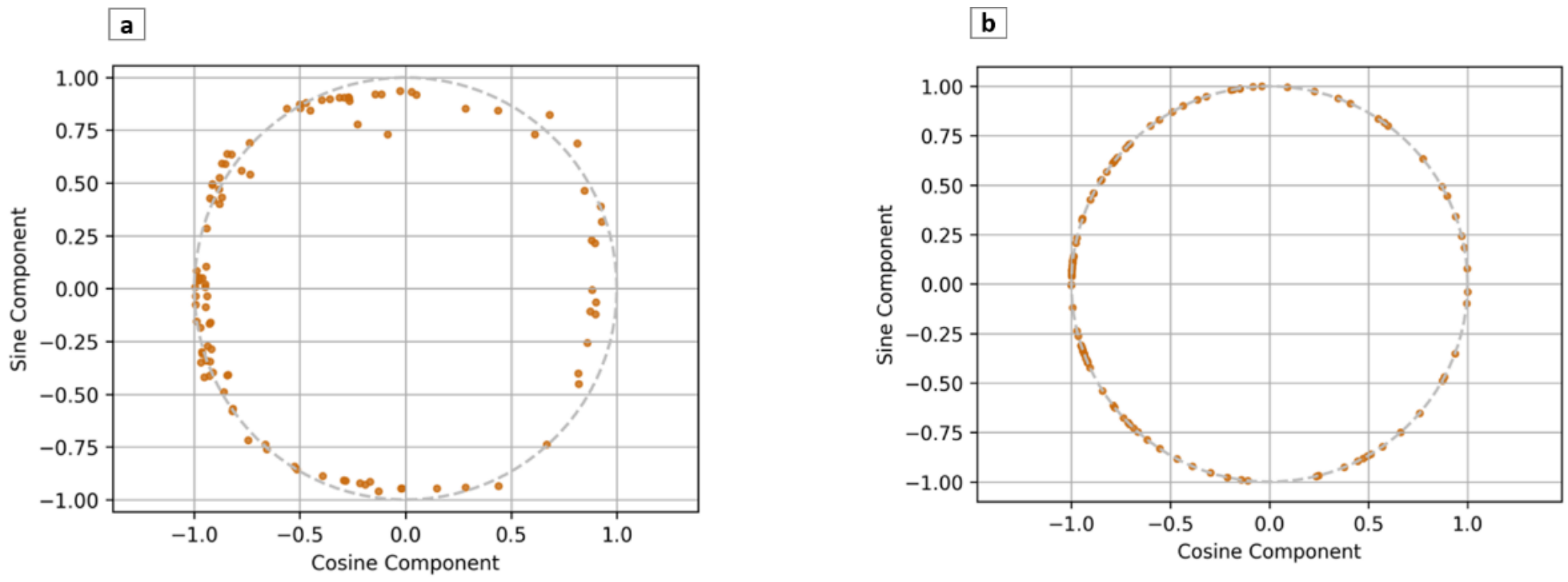

Figure VI: Comparison of phase component distributions for neural networks: (a) without and (b) with Euclidean normalization. Dashed circle indicates $sin^2\varphi + cos^2\varphi = 1$.

## 4. Conclusions

We have presented a physics-constrained neural network that effectively predicts the amplitude and phase of transmitted light from metasurface, demonstrating strong performance across datasets of different sizes (580 and 1075 data points). A sine-cosine representation for phase values and Euclidean normalization within the network architecture enable our model to handle the phase wrapping discontinuity problem and learn essential mathematical and physical constraints. The network achieves accurate prediction despite the small size of imbalanced datasets and the challenging behavior of phase, with consistent $R^2$ values 0.928 for Dataset A and 0.906 for Dataset B. It achieves a mean angular error of 0.114 radians (approximately 6.5°) for Dataset A and 0.124 radians (approximately 7.1°) for Dataset B. Dataset C with 290 data points serves as an extreme test of data scarcity, confirming the expected relationship between dataset size and model performance while showing that the physics-constrained network can still provide predictive capability, albeit with increased volatility in predictions. A key strength of this network is its ability to achieve accurate prediction while preserving physical interpretability. Unlike black-box models, the physics-constrained network ensures that the network's predictions respect fundamental electromagnetic principles like energy conservation and periodic nature of phase. By incorporating physical constraints that narrow the solution space to physically meaningful predictions, the model effectively generalizes small, imbalanced datasets. This approach is extendable to other photonic structures and could be modified to incorporate additional physical constraints for different applications.

**Acknowledgment**

The simulation results used as input datasets in this work were produced using JCMsuite solver, and the authors thank Dr. Sven Burger for the license use. This study was funded by the National Recovery and Resilience Plan (NRRP) – PRIN 2022 PNRR, project STARTrAP – CUP: B53D23024250001 – Grant Assignment Decree No. P2022LB47E_002.

**Data availability**

The datasets used in this study are available from the corresponding author on reasonable request.